# Simulation of Temperature Distribution In a Rectangular Cavity Using Finite Element Method

***Christian Fredy Naa, Suprijadi***
*Theoretical, High Energy and Instrumentation Division, Faculty of Mathematics and Natural Sciences, Institut Teknologi Bandung*
*Jl. Ganesha No. 10 Bandung, Indonesia 40132*
*e-mail address: chris_mail@yahoo.com*

## *Abstract*
*This paper presents the study and implementation of finite element method to find the temperature distribution in a rectangular cavity which contains a fluid substance. The fluid motion is driven by a sudden temperature difference applied to two opposite side walls of the cavity. The remaining walls were considered adiabatic. Fluid properties were assumed incompressible. The problem has been approached by two-dimensional transient conduction which applied on the heated sidewall and one-dimensional steady state convection-diffusion equation which applied inside the cavity. The parameters which investigated are time and velocity. These parameters were computed together with boundary conditions which result in temperature distribution in the cavity. The implementation of finite element method was resulted in algebraic equation which is in vector and matrix form. Therefore, MATLAB programs used to solve this algebraic equation. The final temperature distribution results were presented in contour map within the region.*

## I. INTRODUCTION

In order to analyze an engineering system, a mathematical model is developed to describe the behavior of the system.[1] The mathematical expression usually consists of differential equations and given conditions. These differential equations are usually very difficult to obtain solution if handled analytically.[1] The alternative way to solve the differential equation is using numerical method.

The numerical method has some advantage, not only it can solved complicated equations which are hardly to solve analytically but also can reduce cost which needed to make experiments. Moreover, numerical method is able to predict the physical phenomena so it can be studied and implemented to make some device.

Considering the advantage of the numerical method, this paper uses it to analyze physics phenomena in heat transfer problem. Heat transfer problem that will be discussed is a temperature distribution over a rectangular cavity which is exists because of the sudden temperature difference applied to two opposite side walls of the cavity.

The numerical method which is used is finite element method. The reason why in this paper chooses the finite element method is because the finite element method is one of the numerical method that has received popularity due to its capability for solving complex structural problem.[2] Moreover, the finite element method provides high computational flexibility and on the other hand facilitates a rigorous mathematical error analysis.[3]

After the mathematical analyses take place, the finite element programming requires software to solve algebraic equations which is vector and matrix manipulation. Therefore, for this paper the programming uses MATLAB interactive software.

## II. PROBLEM DEFINITION

The objective of this paper is finding the temperature distribution in a rectangular cavity as shown in Figure 1.

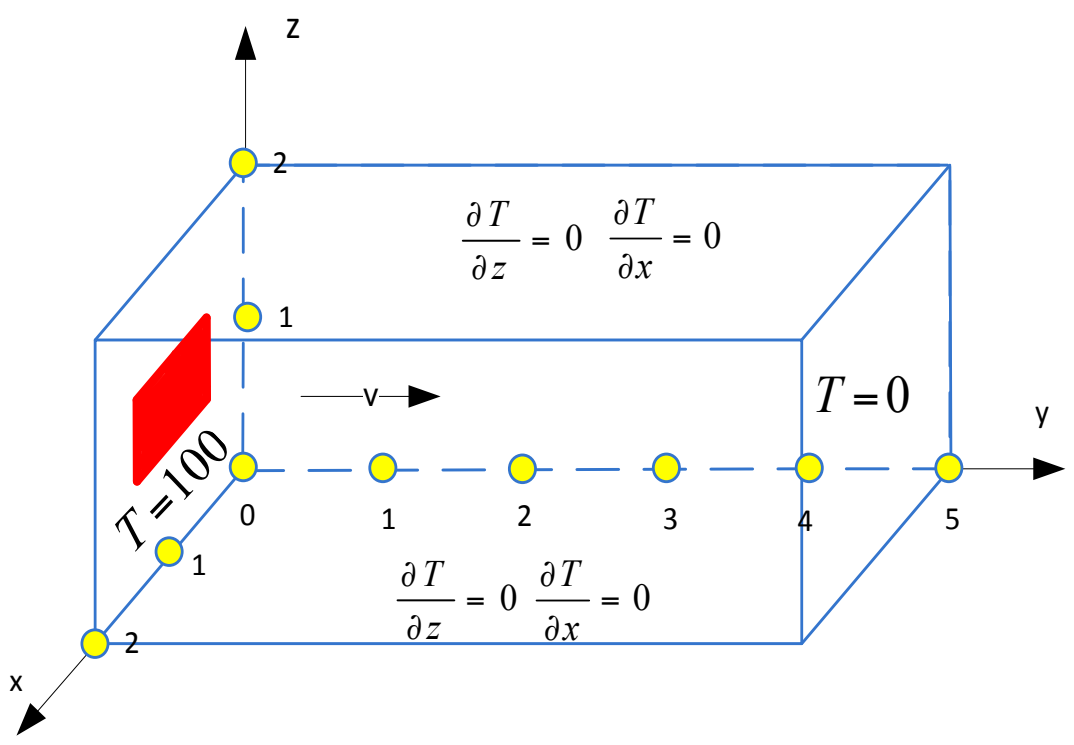

**Figure 1.** Rectangular Cavity with Boundary Condition

The cavity considered built by thin metal sheet homogenous material properties as the boundaries, inside the cavity consist a homogenous fluid substance. The initial temperature inside the cavity is considered to be $0^{o}C$. Suddenly, the cavity heated on the left sidewall while the right one is keeps at zero temperature, meanwhile the other side of the cavity is remain insulated.

The heated sidewall (on $x$ and $z$ axis) is shown in Fig. 2.

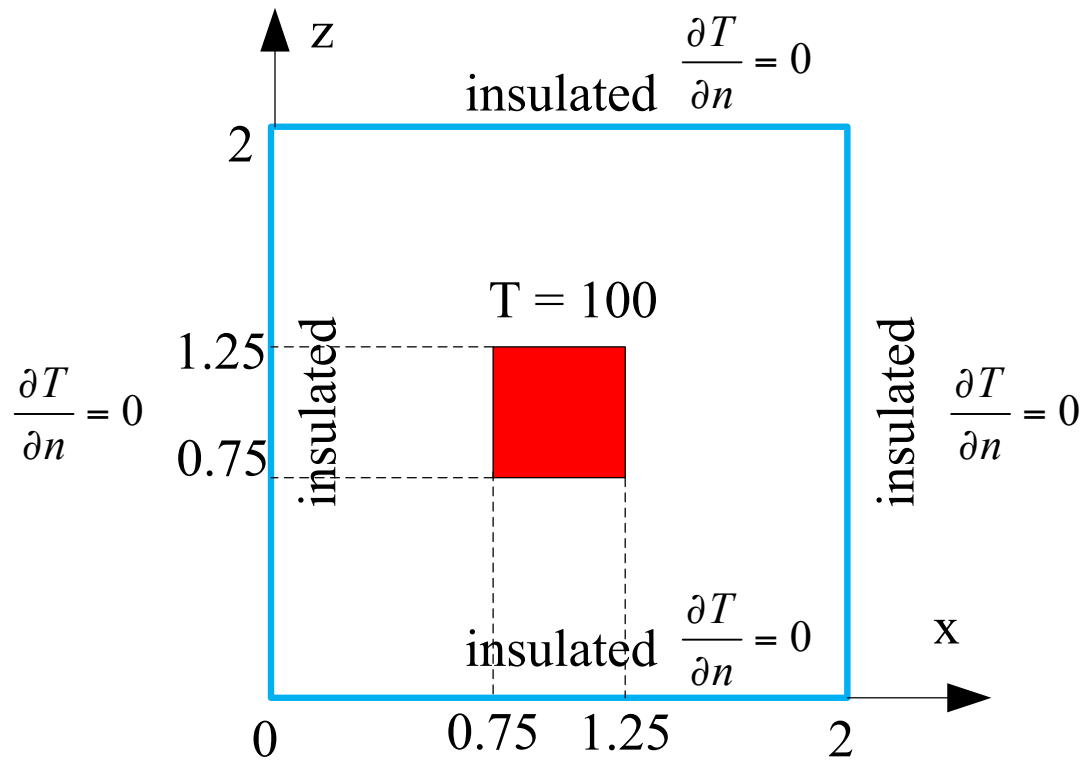

**Figure 2.** Left Sidewall of the Cavity with Boundary Conditions

It will be considered as time dependent conduction heat transfer problem. The equation which governs this problem is conduction (diffusion) equation.[4]

$$\frac{\partial^2 T}{\partial x^2} + \frac{\partial^2 T}{\partial z^2} = \frac{1}{\alpha}\frac{\partial T}{\partial t} \quad (1)$$

Here, α is material characteristic constant, its measure of how quickly a material can carry heat away from hot source. For simplicity, take $\alpha = 1$. The boundary conditions for this left side of the cavity are

1. $\frac{\partial T(x,0)}{\partial z} = \frac{\partial T(x,2)}{\partial z} = \frac{\partial T(2,z)}{\partial x} = \frac{\partial T(0,z)}{\partial x} = 0,$
2. $T = 100$ for $0.75 < x < 1.25$ and $0.75 < z < 1.25$

Next, the fluid inside the cavity is considered homogenous and incompressible (constant density and pressure) which flow by the present of constant velocity $v$ in $y$ axis direction (one-dimensional direction). This yields to a laminar flow inside the cavity. This means the flow inside the cavity is concerned to be a natural convective flow. In this paper, the analysis of the flow is considered in steady state condition or time independent. The equation which governs this problem is convection-diffusion equation which is similar with diffusion equation except the present of the convective terms.[5]

$$\frac{1}{\alpha}\frac{d^2 T}{dy^2} = v\frac{dT}{dy} \quad (2)$$

Similar with thermal diffusivity $\alpha$, the velocity $v$ explained how quickly fluid can carry heat away from hot source.

# III. FINITE ELEMENT METHOD

The basic idea in the finite element method is to find the solution of a complicated problem by replacing it by a simpler one.[6] Since the actual problem is replaced by a simpler one; the solution is only an approximation rather than the exact one. In the finite element method the solution domain is divided into small domain called *elements*; these elements connected each other by *nodes*, therefore this method called finite element. This subdivision process is called discretization.[6]

In each element, a convenient approximate solution is assumed and the conditions of overall equilibrium of the structure are derived.[6] The derivation is using several mathematical analyses, likewise integration by parts, weighted residual and Galerkin's method. Once this analyses done, each element will form matrix equation, which resulting in global matrix equations. These global matrix equations solved by MATLAB to get the solution for each node.

All this processes are the basic procedures for the finite element method. Systematically, each of the basic procedures will be discussed separately for conduction and convection problem.

## Finite Element Method for Conduction Problem

Figure 3 shows the discretization process for the conduction on left side of the cavity. The region is considered as 128 triangular elements (red numbers) and 81 interconnected nodes (blue numbers).

After the discretization process, the mathematical analyses take place; it does begin with method of weighted residual. Weighted Residual method is used for finding the approximation for the governing equation. The method is multiply the trial function and a weight function, then we take the integral value in all domains, the value must be 0 (this value means, we make the error for the assumption function equal to zero or no error at all).

$$I = \int_{\Omega} (wT)\,d\Omega = 0, \tag{3}$$

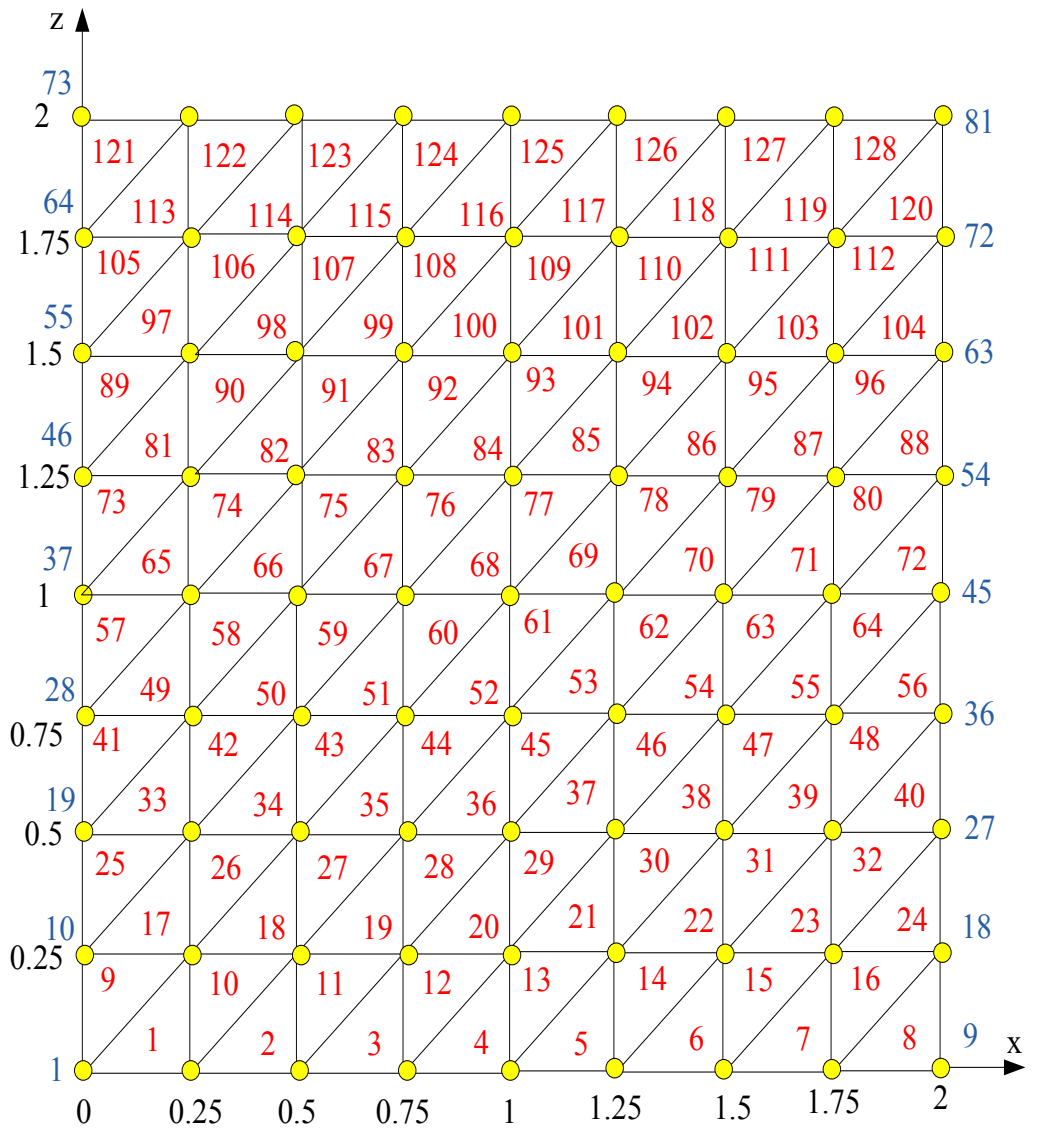

**Figure 3.** Discretization Process of Heat Conduction Problem on the left side of the cavity

Here, $T$ is trial equation and $w$ is weight function, by using Eq. (3) and integration by parts, the governing equation on Eq. (1) becomes,

$$I = \int_{\Omega} w\frac{\partial T}{\partial t}\,d\Omega + \int_{\Omega}\left(\frac{\partial w}{\partial x}\frac{\partial T}{\partial x} + \frac{\partial w}{\partial z}\frac{\partial T}{\partial z}\right)d\Omega - \int_{\Gamma n} w\frac{\partial T}{\partial n}\,d\Gamma = 0 \tag{4}$$

Here, $\Omega$ is all domain and $\Gamma$ is boundary condition. The integration result in another term, the boundary term. When the boundary is insulated so there is no flux $\left(\frac{\partial T}{\partial n} = 0\right)$, the boundary term vanish.

The $T$ function or trial function must be simple and easily to differentiate and integrated, the most useful function is polynomial function because this function can be easily differentiate and integrated. It also has advantages; if necessary the approximation quality can be improved by increase the polynomial degree. The $w$ function is take as same as $T$ function, this technique called Galerkin's Method.[1]

From Fig. 3, the element is triangular element and it shown in Fig. 4. The triangular element has three nodes at the vertices of the triangle and the variable interpolation within the element is linear in $x, z$, and $t$. The interpolation for linear triangular element becomes

$$T(x,z,t) = H_1(x,z)T_1(t) + H_2(x,z)T_2(t) + H_3(x,z)T_3(t) \tag{5}$$

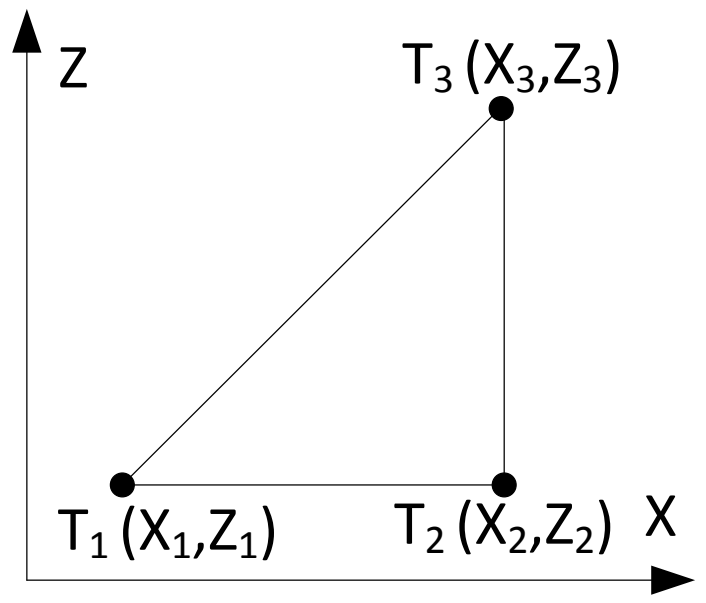

**Figure 4.** Triangular Element

Here, $H_i(x,y)$ are the shape functions,

$$H_1(x,z) = \frac{1}{2A}[(x_2z_3 - x_3z_2 + (z_2 - z_3)x + (x_3 - x_2)z] \tag{6a}$$

$$H_2(x,z) = \frac{1}{2A}[(x_3z_1 - x_1z_3 + (z_3 - z_1)x + (x_1 - x_3)z] \tag{6b}$$

$$H_3(x,z) = \frac{1}{2A}[(x_1z_2 - x_2z_1 + (z_1 - z_2)x + (x_2 - x_1)z] \tag{6c}$$

Here, $A$ is the area of the triangle.

Equation (4) consists of two terms, the spacial terms and transient terms. The spacial terms with substitution of the shape function together with Galerkin's Method becomes

$$[K^e]\{T^e\} = \int_{\Omega^e}\left(\frac{\partial w}{\partial x}\frac{\partial T}{\partial x} + \frac{\partial w}{\partial z}\frac{\partial T}{\partial z}\right)d\Omega = \int_{\Omega^e}\left(\begin{Bmatrix}\frac{\partial H_1}{\partial x}\\ \frac{\partial H_2}{\partial x}\\ \frac{\partial H_3}{\partial x}\end{Bmatrix}\begin{bmatrix}\frac{\partial H_1}{\partial x} & \frac{\partial H_2}{\partial x} & \frac{\partial H_3}{\partial x}\end{bmatrix} + \begin{Bmatrix}\frac{\partial H_1}{\partial z}\\ \frac{\partial H_2}{\partial z}\\ \frac{\partial H_3}{\partial z}\end{Bmatrix}\begin{bmatrix}\frac{\partial H_1}{\partial z} & \frac{\partial H_2}{\partial z} & \frac{\partial H_3}{\partial z}\end{bmatrix}\right)d\Omega \begin{Bmatrix}T_1\\ T_2\\ T_3\end{Bmatrix}^e \tag{7}$$

Here, $\Omega^e$ is the element domain. Performing integration after substituting the shape functions will gives symmetric matrix.

Meanwhile element matrix for transient term becomes

$$[M^e]\{\dot{T}^e\} = \int_\Omega w\frac{\partial T}{\partial t}d\Omega =$$

$$\int_\Omega \begin{Bmatrix}H_1\\H_2\\H_3\end{Bmatrix}[H_1 \quad H_2 \quad H_3]d\Omega\begin{Bmatrix}\dot{T}_1\\\dot{T}_2\\\dot{T}_3\end{Bmatrix} \tag{8}$$

Here, $\dot{T}$ is temperature derivative with respect to time, computation Eq. (8) results in

$$[M^e] = \frac{A}{12}\begin{bmatrix}2 & 1 & 1\\1 & 2 & 1\\1 & 1 & 2\end{bmatrix} \tag{9}$$

Once again, this matrix form result in symmetrical matrix. Therefore, the final matrix equation for Eq. (4) becomes

$$[M]\{\dot{T}\}^t + [K]\{T\}^t = \{F\}^t \tag{10}$$

The column vector $\{F\}^t$ is the boundary conditions to satisfy the matrix equation. Because this equation should be true for any time, the superscript $t$ in Eq. (10) placed to denote the time when the equation is satisfied. Furthermore, matrices $[M]$ and $[K]$ are independent of time.
To solve Eq. (10), the finite difference method takes place. Equation (10) can be written at time $t + \Delta t$

$$[M][\dot{T}]^{t+\Delta t} + [K][T]^{t+\Delta t} = \{F\}^t \tag{11}$$

The time derivatives in the backward difference is[1]

$$\{\dot{T}\}^{t+\Delta t} = \frac{\{T\}^{t+\Delta t} - \{T\}^t}{\Delta t} \tag{12}$$

Use Eq. (11) with Eq. (10) results in

$$([M] + \Delta t[K])\{T\}^{t+\Delta t} = \Delta t\{F\}^{t+\Delta t} + [M][T]^t \tag{13}$$

By choosing the appropriate step time and start with $t = 0$, the final solution will be found. All these calculation implemented with MATLAB structured program.

## Finite Element Method for Convection Problem

Meanwhile, the discretization process for convection inside the cavity is shown in Fig. 5. The domain considers build up by 9 equal size small elements and interconnected by 10 nodes.

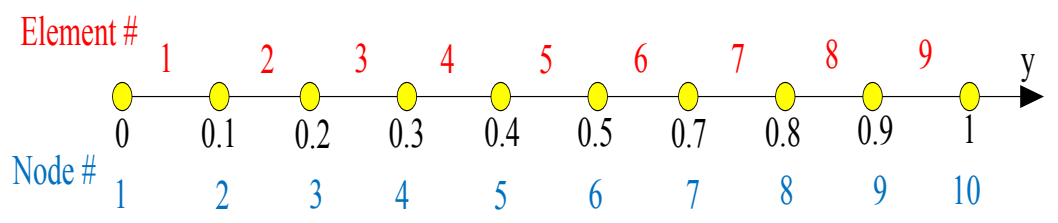

**Figure 5.** Discretization Process for Heat Convection-Diffusion Problem

Figure 5 shows representation from the actual problem. The actual problem consist 5 unit lengths on y axis, this means the actual problem consist 45 elements with 50 nodes.

After the discretization, similar with conduction problem, the method of weighted residual and Galerkin's takes place, Eq. (2) becomes

$$I = \int_\Omega \left(w\frac{d^2T}{dy^2} - vw\frac{dT}{dy}\right)d\Omega = 0 \tag{14}$$

Here, $\Omega$ is the element domain. Thus, by Integration by parts[6] Eq. (14) becomes

$$I = \int_\Omega \left(-\frac{dw}{dy}\frac{dT}{dy} - vw\frac{dT}{dy}\right)d\Omega + \left[w\frac{dT}{dy}\right]_\Omega = 0 \tag{15}$$

Equation (15) contain extra boundary terms which neglected due to weight function in boundary is zero. The interpolation for one-dimensional element is

$$T = H_1(y)T_i + H_2(y)T_{i+1} \tag{16}$$

Where

$$H_1(y) = \frac{y_{i+1} - y}{h_i} \tag{17a}$$

$$H_2(y) = \frac{y - y_i}{h_i} \tag{17b}$$

$$h_i = y_{i+1} - y_i \tag{17c}$$

Subscript $i$ denotes node number and $h_i$ is element length. By using the interpolation formula in Eq. (16) and the Galerkin's method, Eq. (15) becomes

$$\int_{y_i}^{y_{i+1}} \left(-\frac{dw}{dy}\frac{dT}{dy} - vw\frac{dT}{dy}\right)dx = \left(-\int_{y_i}^{y_{i+1}}\left(\begin{Bmatrix}H'_1\\H'_2\end{Bmatrix}[H'_1 \quad H'_2] - v\begin{Bmatrix}H_1\\H_2\end{Bmatrix}[H'_1 \quad H'_2]\right)dy\right)\begin{Bmatrix}T_i\\T_{i+1}\end{Bmatrix} \tag{18}$$

where ()' denotes the derivatives with respect to $x$. Evaluating Eq. (18) gives

$$[K^e]\{T^e\} = \left(-\frac{1}{h_i}\begin{bmatrix}1 & -1\\-1 & 1\end{bmatrix} - \frac{v}{2}\begin{bmatrix}-1 & 1\\-1 & 1\end{bmatrix}\right)\begin{Bmatrix}T_i\\T_{i+1}\end{Bmatrix} \tag{19}$$

The matrix final form becomes

$$[K]\{T\} = \{F\} \quad (20)$$

Similar with conduction analysis, the column vector $\{F\}$ is the boundary conditions to satisfy the matrix equation. All these calculation implemented with MATLAB structured program.

# IV. THE APPROACHING MODEL

To solve this problem, the inside of the cavity was divided into seven layers, each layer consist of seven lines, but not include the boundaries. That means the three dimension problem was approached by two dimension model and then by one dimension model. This approaching model is shown in Fig. 6.

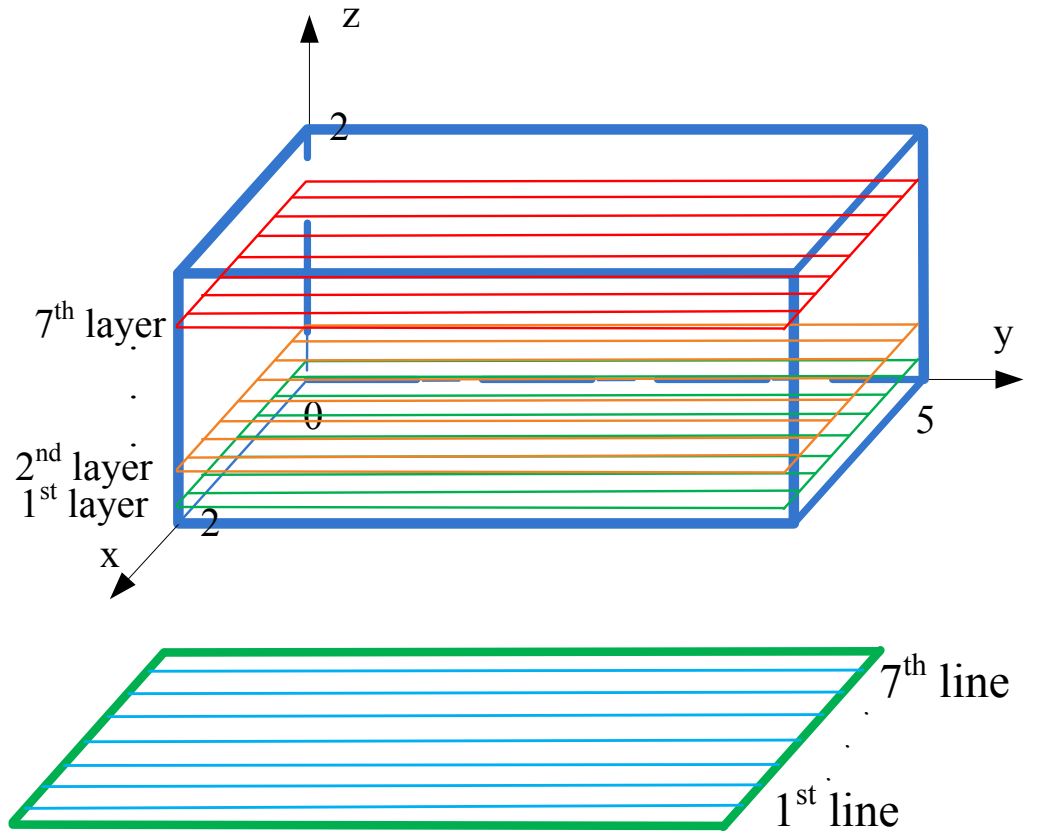

**Figure 6.** The Approaching Model: The Cavity Divided Into Seven Layers, Each Layer Consist of Seven Lines

To get the temperature distribution inside the cavity several steps took place. First, temperature distribution on the left side of the cavity solved using the conduction finite element formulation which yields to the temperature value in each node on the left side of the cavity. Second, these values became the boundary conditions for convection-diffusion finite element formulation which yields to the temperature distribution on each line. Next, these line temperature distributions which in the same layer were combined to get the contour map for each layer. These contour maps were the final solution which will be given in the next sections.

# V. RESULT AND DISCUSSION

The simulations of the temperature distribution inside the cavity involve two parameters, the time $t$ and the velocity $v$. The time parameter governs the left side heat conduction. Meanwhile, the velocity parameter governs the heat convection-diffusion. Therefore, both parameters were discussed separately.

## Time Based

The time based analysis took the comparison between time at 0.8 second and 10 second with the same value of velocity ($1\ m/s$). This comparison taken because at the time 10 second, the conduction at the left side of the cavity has reach steady state condition. The results are shown in Fig. 7 and Fig. 8. The time parameter which has been plotted is 0.8 second and 10 second gives different result in heat conduction on left side of the cavity. At 0.8 second, the heat has just begun to spread from the middle of the metal sheet, this yields to different boundary condition to the convection-diffusion heat mode inside the cavity. As a result, each convection-diffusion line on the layers gave different result. Meanwhile, at 10 second, the conduction on the left side of the cavity has reach steady state condition which the temperature reaches $100^{o}C$ in all domain of the metal sheet. This yields to same boundary condition to the convection-diffusion heat modes inside the cavity. This means, in all convection-diffusion line gave the same result on the layers.

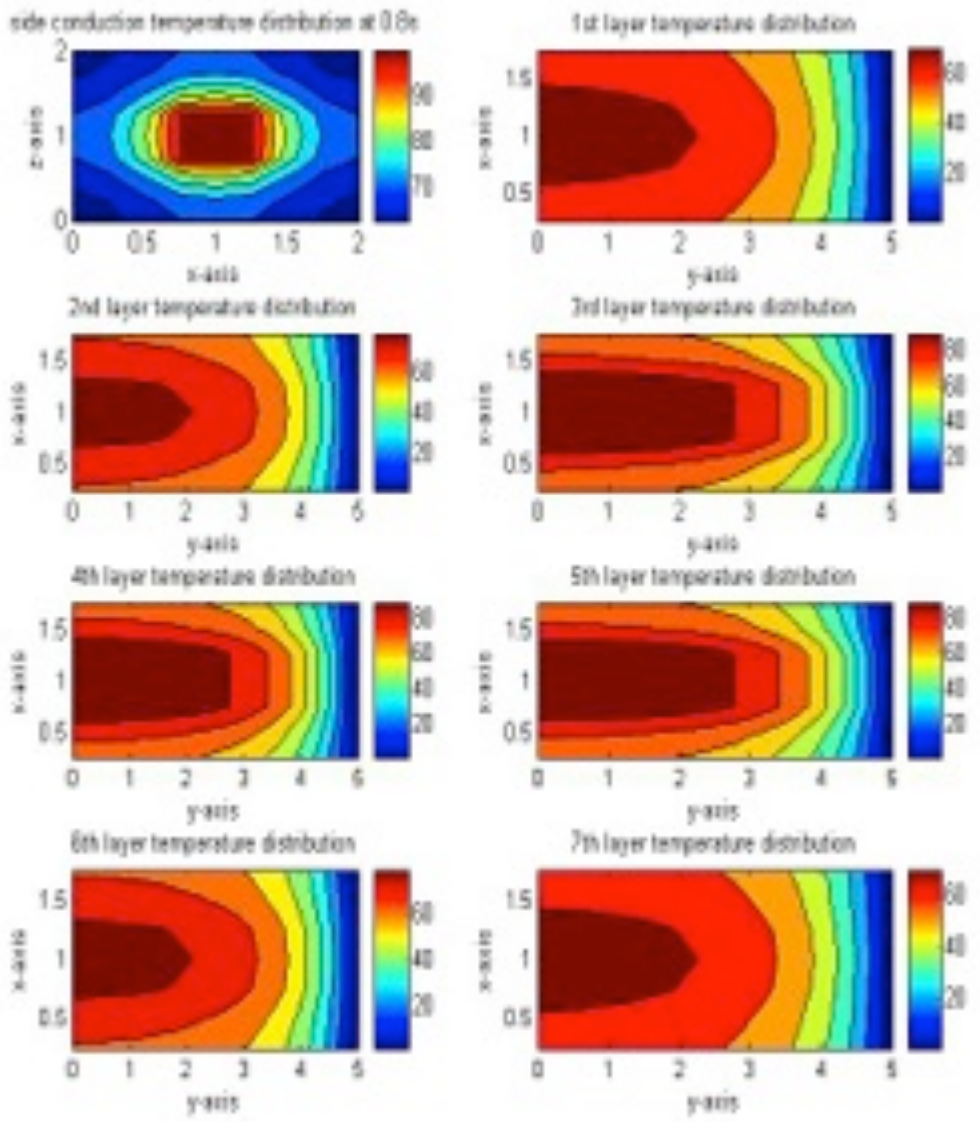

**Figure 7.** Temperature Distributions at 0.8 second and Velocity 1 m/s

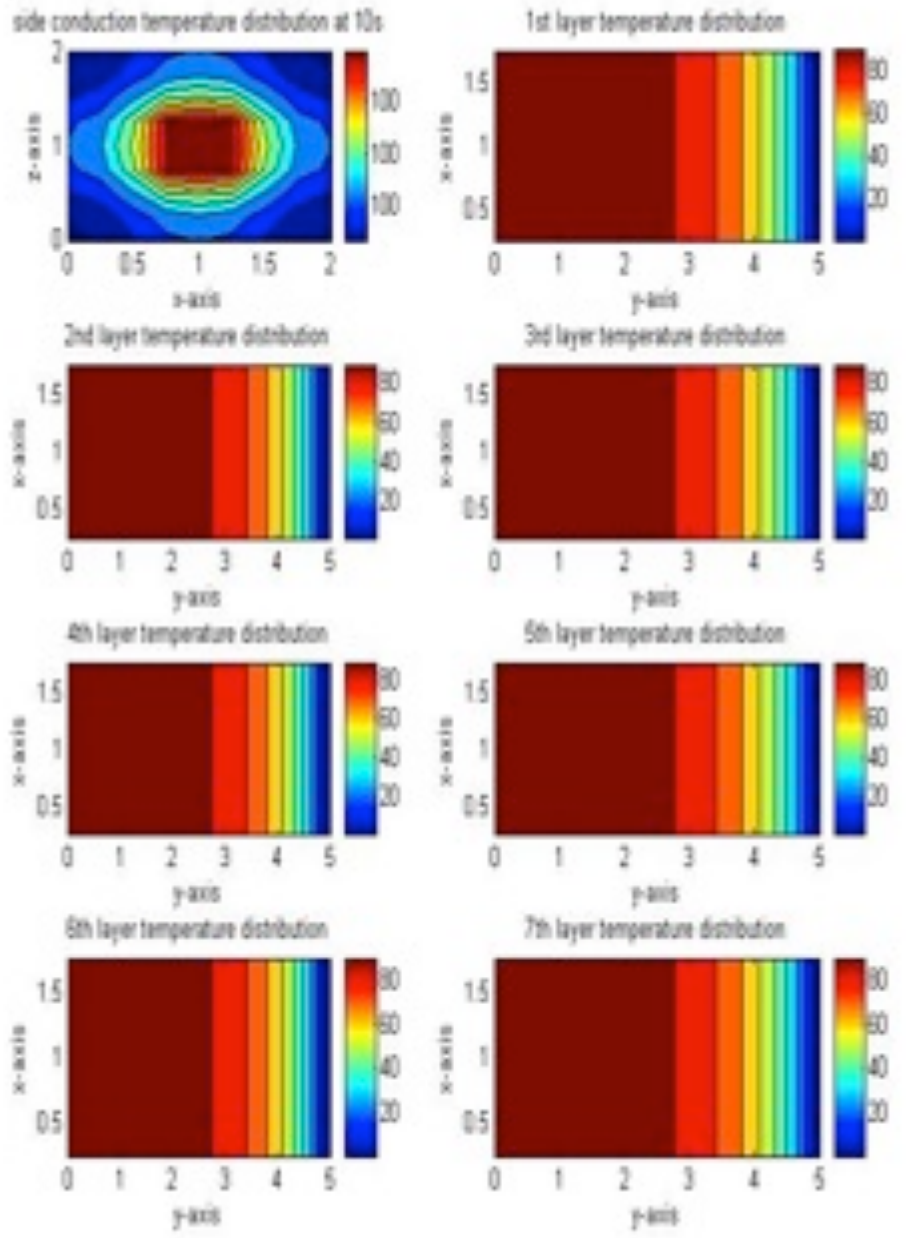

**Figure 8.** Temperature Distributions at 10 second and Velocity 1 m/s

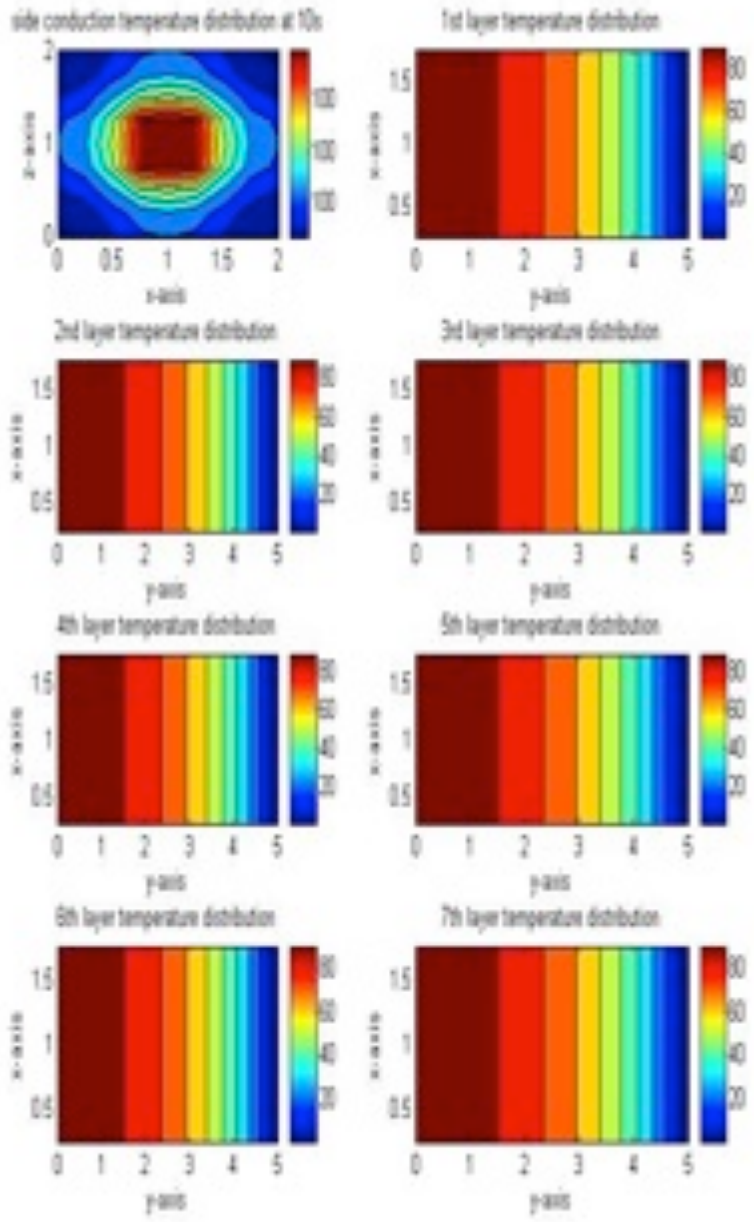

**Figure 9.** Temperature Distributions at 10 second and Velocity 0.5 m/s

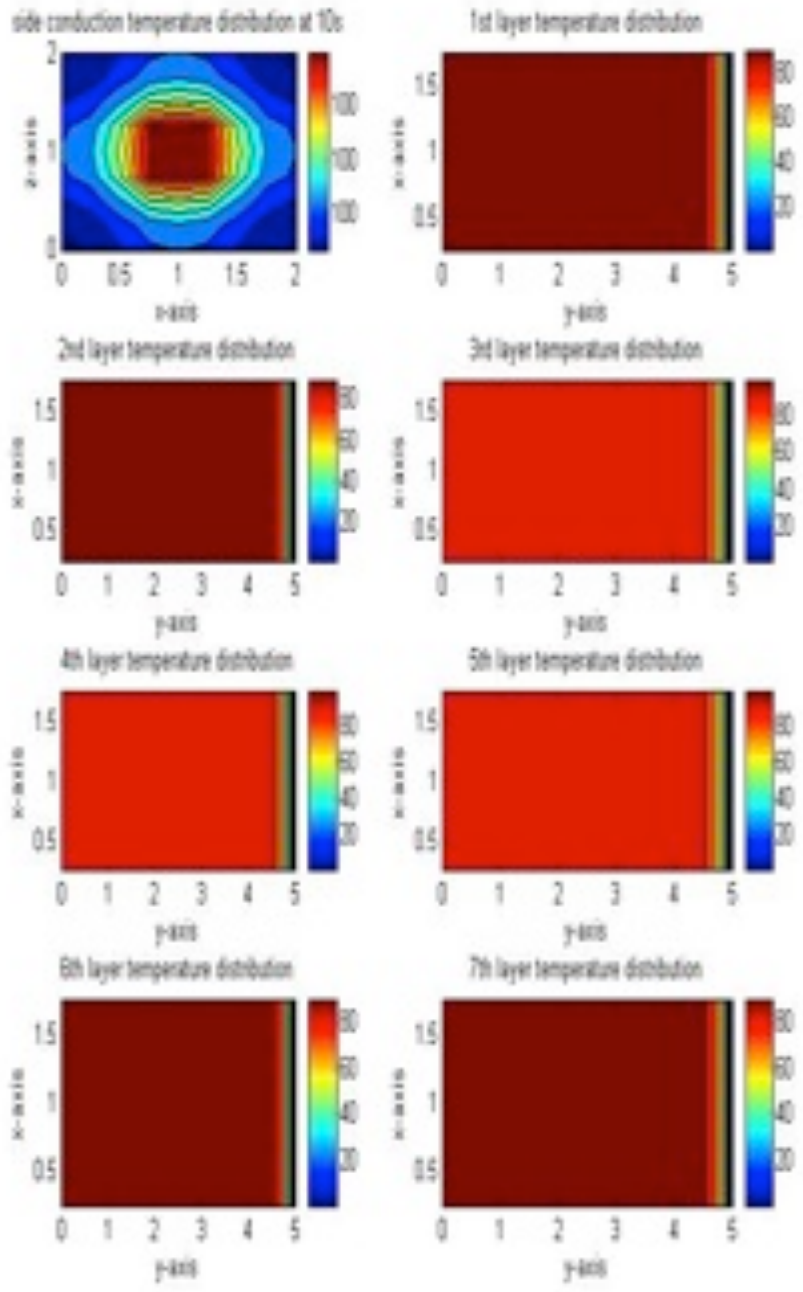

**Figure 10.** Temperature Distributions at 10 second and Velocity 5 m/s

## Velocity Based

The velocity based analysis took the comparison between velocity $0.5\ m/s$ and $5\ m/s$ at the same time ($10\ s$). The results are shown in Fig. 9 and Fig. 10. This velocity differences made the heat flow faster as the velocity increase. Therefore, when the velocity has low value, the heat distribution inside the cavity is quite even, means that the high temperature and low temperature was spread evenly. As the velocity higher, the heat distribution inside the cavity was not spread evenly anymore. Therefore, the high temperature dominated the low one.

## VI. CONCLUSION

This paper presents the study and implementation of finite element method to find the temperature distribution inside a rectangular cavity. The time and velocity parameters that influence the temperature distribution were analyzed by the help of the contour map.

It was shown that the time parameter governs the conduction on the heated sidewall, this implication is due to the conduction is time dependent problem (Eq. 1). As the time increase, the overall temperature in the sidewall becomes increase too and finally it reached steady state condition after 10 second (Fig. 9 and Fig. 10). The results of the heated sidewall conduction yield to the boundary condition for the convection-diffusion inside the cavity. It was shown in contour map for each layer inside of the cavity.

Meanwhile, the velocity parameter governs the convection-diffusion inside the cavity. This implication is due to the convective terms in convection-diffusion steady state equation (Eq. 2). As the velocity increase, the temperature distribution inside the cavity was dominated by the temperature conditions on the heated sidewall. These findings were shown in contour map layers (Fig. 9 and Fig.10).